\documentclass[12pt]{article}
\usepackage{mathtools}
\usepackage[utf8]{inputenc}
\usepackage{amsmath}
\usepackage[margin=01.0in]{geometry}
\usepackage{amssymb}
\usepackage{amsfonts}
\usepackage{amsthm}
\usepackage{enumitem}
\usepackage{xcolor}
\usepackage{lipsum}
\usepackage{setspace} 
\numberwithin{equation}{section}

\let\emptyset\varnothing

\usepackage{tabularx}

\usepackage{graphicx}
\graphicspath{ {./images/} }

\theoremstyle{plain}

\theoremstyle{definition}

\theoremstyle{remark}

\numberwithin{thm}{section}

\title{The Market for Lemons and the \\ Regulator's Signalling Problem}
\author{Roy Long \thanks{University of Chicago Economics \\ Original version July 2022. This version with major revisions December 2023}}
\date{December 2023}
\begin{document}

\maketitle

\begin{abstract}
The Market for Lemons is a classic model of asymmetric information first studied by Nobel Prize economist George Akerlof. It shows that information asymmetry between the seller and buyer may result in market collapse or some sellers leaving the market. ``Lemons" in the used car market are cars of poor quality. The information asymmetry present is that the buyer is uncertain of the cars' true quality. I first offer a simple baseline model that illustrates the market collapse, and then examine what happens when regulation, ie. a DMV is introduced to reveal (signal) the true car quality to the buyer. The effect on the market varies based on the assumptions about the regulator. The central focus is on the DMV's signal structure, which can have interesting effects on the market and the information asymmetry. I show that surprisingly, when the DMV actually decreases the quality of their signal in a well constructed way, it can substantially increase their profit. On the other hand, this negatively effects overall welfare.
\end{abstract}

\newpage
\section{Baseline Model}

In this article, we will review the famous Market for Lemons by Akerlof. It is a simple model that shows how information asymmetry between buyer and seller can have very interesting effects on a market. Then we will introduce a certain third party and study the implications of information design.

``Lemons" in a market are essentially bad goods of low quality. Akerlof showed that information asymmetries in the market can cause the quality of goods to degrade until only ``lemons" are left.

We use a car market for our setting as in Akerlof's model. The information asymmetry here is that the quality of the car is known to the sellers (car dealers) but not to buyers. We can represent the quality of a used car by a hidden type $\theta \in [0,1],$ where $\theta = 0$ means a busted car and $\theta = 1$ means in new condition.

Payoffs for the buyer and seller are given by:
\\
\\
\begin{tabular}[t]{|l|cc|c|}
\multicolumn{3}{c}{Market for lemons}\\\hline
&\ no sale &\ sale \\\hline
seller& $\pi \theta$ & $p$ \\
buyer& $0$ & $\theta - p$ \\\hline
\end{tabular}
\\
\\
\\ where $\pi < 1$ indicates that the seller is somewhat tired of the car and it has less utility for them (especially the longer they're unable to sell the car), though $\pi$ is not too small, so we assume $\pi > \frac{1}{2}.$ And $p$ is the price of the car determined endogenously by the market.

Now each seller has a type represented by their $\theta \in [0,1].$ Suppose all sellers are on the market. Then we note that the buyer accepts $p$ only if $p \leq E(\theta) = \frac{1}{2}$ under uniform prior. But then given $p = \frac{1}{2},$ seller gets net gain of $p - \pi \theta = \frac{1}{2}  - \pi \theta,$ so she'll only sell if $\theta \leq \frac{1}{2\pi} < 1.$

Now since only the sellers with $\theta \leq \frac{1}{2\pi}$ will sell, the buyer will only buy if $p \leq E \left(\theta \mid \theta \leq \frac{1}{2\pi} \right) = \frac{1}{4\pi}.$ But by similar reasoning then the seller will only sell if $\frac{1}{4 \pi} - \pi \theta \geq 0,$ or $\theta \leq \frac{1}{4 \pi^2} < \frac{1}{2\pi}.$ 

This reasoning can easily be repeated recursively, and it follows by induction at the $n$-th step that the only car dealers left willing to sell are $\theta \leq \frac{1}{(2\pi)^n}$ and buyers will only buy if $\theta \leq \frac{1}{2(2\pi)^n}$ but clearly as $n \to \infty$ the $\frac{1}{(2\pi)^n} \to 0,$ so in fact only $\theta = 0,$ the lemons will sell. Do we have a market anymore? Apparently not. 

This kind of reasoning is called unraveling to the bottom, or retrograde analysis or working backwards. When this was first proposed, it had profound implications, because it showed that under reasonable assumptions and information asymmetry, a market could suffer market failure. 

What is a proposed solution? There are indeed market solutions on both supply and demand side. On the supply side, sellers may provide a warranty to assuage buyers' concerns about quality uncertainty, but this may be costly for sellers with low quality cars. The sellers may also increase the proportion of higher quality cars, so the buyer's prior belief of buying a good car is higher. However this may not be so easy to achieve for a particular seller type. On the demand side, buyers may seek third party quality specialists to examine the cars, thereby removing the information asymmetry. We consider a third alternative, a regulation based solution to create a different kind of incentive in the market.

Let us assume the DMV, a regulator, will assist the buyer in determining the true value of the seller's car quality, thereby mitigating the effects of the information asymmetry.

Suppose the DMV offers certificates that disclose quality for free. (We will re-evaluate this assumption later.)
Each seller then chooses whether to disclose $\theta$ or not.
Thus, the buyer is willing to pay for
\begin{itemize}
  \item $w = \theta$ if $\theta$ is disclosed, and 
  \item $w = E(\theta \mid \text{undisclosed}) = \widehat{\theta}$ if $\theta$ is not disclosed.
\end{itemize}

However, note that given $\widehat{\theta},$ the seller types $\theta \geq \widehat{\theta}$ will disclose, and sellers with $\theta < \widehat{\theta}$ are better off not disclosing.  But this means that $\widehat{\theta} = E(\theta \mid \text{undisclosed}) = E(\theta \mid \theta < \widehat{\theta}) = \frac{\widehat{\theta}}{2},$ which solves $\widehat{\theta} = 0.$ Thus, only the lemons won't disclose.

We can easily understand this substantively with a similar recursive argument. Suppose some sellers $\theta$ disclose and some don't disclose. Consider a nondisclosing seller with $\theta$ on the high end of the types that don't disclose. To the buyer, her type can only be evaluated as best as $E(\theta \mid \text{undisclose}),$ which is strictly lower than her true type $\theta$. In this case the seller is essentially being lumped in with the low $\theta$'s that are also not disclosing, and that is not good for her. So that means actually the seller should disclose. So then more types will disclose, and we can keep repeating this argument until all but the $\theta = 0$ will disclose, because as long as multiple types of $\theta$ are not disclosing then some $\theta$ in that group is being disadvantaged by the information asymmetry and the rest of the nondisclosing group.

Rather remarkably the mere introduction of a DMV completely reverses the market failure from no one willing to sell except the lemons to everyone willing to sell except the lemons.

This suggests that regulation, ie. a DMV, may solve the problem. However, the DMV probably won't give the seller $\theta$ for free because it probably costs them money to do an inspection etc. Or the DMV also wants to make a profit.

Suppose the DMV charges a fee of $c > 0$ for certificates. Then each seller chooses whether to buy a certificate. The certificate will display the car's true quality $\theta$ to the buyer. Sellers that don't buy the certificate will still have their $\theta$ concealed.
However, an important implication of this is that the buyer is aware that some sellers on the market may disclose and others won't and can reason about this when evaluating a seller.

This means that the buyer is willing to pay for 
\begin{itemize}
  \item $w = \theta$ if $\theta$ is disclosed and 
  \item $w = E(\theta \mid \text{undisclosed}) = \widehat{\theta}$ if $\theta$ is not disclosed.
\end{itemize}

Given $\widehat{\theta},$ the seller is better off disclosing if $$\theta - c \geq \widehat{\theta}$$ and better off not disclosing if $$\theta - c < \widehat{\theta}.$$ Thus, $$\widehat{\theta} = E(\theta \mid \text{undisclosed} ) = E(\theta \mid \theta < \widehat{\theta} + c) = \frac{\min\left(\widehat{\theta} + c, 1\right)}{2},$$ which solves $2 \widehat{\theta} - \min \left( \widehat{\theta} + c, 1 \right) = 0$ ie. $\min \left(\widehat{\theta} - c, 1 - 2 \widehat{\theta} \right) = 0,$ which means $\widehat{\theta} = \min \left( c, \frac{1}{2} \right).$
If $c > \frac{1}{2},$ then $\widehat{\theta} = \frac{1}{2}$ and $\theta - c < 1 - c < \frac{1}{2} = \widehat{\theta},$ so no one will disclose.
If $c \leq \frac{1}{2},$ then $\widehat{\theta} = c,$ and the ones that disclose are $\theta \geq c + c = 2c.$

Given this, the DMV maximizes their profit of $P(c) := c(1-2c)$ for $c \leq \frac{1}{2}.$ This is maximized at $c = \frac{1}{4},$ with DMV's profit being $\frac{1}{8}$ and only car dealers with $\theta \geq \frac{1}{2}$ disclose.

We see here the DMV is maximizing under a tradeoff. The DMV wants more sellers to buy the certificate, but by making the certificate's cost higher, fewer sellers will buy it. So by charging the fee for the certificate, the DMV is not able to ensure all types will disclose.

\newpage

\section{Regulation with Noisy Signalling}

We now turn to our main focus, the DMV's signalling problem. Can the DMV cheat? What if the DMV gives the seller a noisy signal? Realistically, the DMV's signal can have some error in it. Problems might arise in the inspection or they might do a poor job at inspecting the car. Suppose the DMV is profit-motivated and seeks to maximize its profit, with welfare being only a secondary concern. It may also be costly for the DMV to do a full inspection of the car to determine its true type $\theta.$ Instead, they might shirk and do a low level inspection, revealing a noisy signal, or maybe they might even try to give the seller a fake signal completely. This is likely not beneficial for social welfare. Can the DMV still prevent the market failure with a lower quality signal and get away with it, achieving a pareto efficient outcome? Can the DMV even improve on their total profit (more than their profit from sending the true $\theta$ signal)? These are some interesting questions that motivate our study of the DMV's signalling problem. \\

We begin by stating several possible information structures and discussing them briefly.  \\
First signal structure: The DMV gives a signal $\theta + \varepsilon$ where $\varepsilon$ is drawn from a distribution with cdf $F$ and density centered at $0.$ This model is realistic, for example, letting $\varepsilon$ be standard normal noise. This signal can be interpreted as the DMV attempting to perform an accurate inspection of the car, but the result may have some error in it. 

Second signal structure: DMV gives $\theta$ with probability $p$ and gives an uninformative signal with probability $1-p,$ possibly just drawn randomly from $[0,1],$ ie. $s \sim U(0, 1).$ This model is interesting as there is again noise in the signal, but unlike the first signal structure, the signal is not centered around $\theta.$ Here the DMV is choosing a lottery over two signal structures - giving a perfect signal ($s = \theta$) and an uninformative signal ($s \sim U(0,1)$ independent of $\theta$). It may not be clear why such a signal structure may be beneficial to the DMV compared to a perfect signal $(p = 1),$ but it turns out that the randomized arbitrary signal in the $s \sim U(0, 1)$ is not as innocuous as upon first glance. Given the particular market structure and how the buyers view different car sellers under the information asymmetry, the inclusion of this noise can create an interesting effect on the market. This signal structure leads to new considerations for sellers in their decision to disclose or not. In particular, lower seller types may have greater incentive to disclose. This signal structure  will be the main topic of this paper.

Third signal structure: DMV gives a signal drawn from some distribution $G$ with support containing $[0,1]$ independent of $\theta.$ This is a totally uninformative signal. Such a signal can be interpreted as the DMV does not perform any inspection of the car at all and completely fabricates an arbitrary quality for the seller to display to the buyer. Notice that this is equivalent to the $p = 0$ signal of the second  As expected, this cannot be sustained. However, to gain intuition about the model and the signal structure, we will begin with this ``completely fake" signal which provides no information about the true car quality and discuss how the market responds to it formally. \\

Let us consider how the market responds to the signal $s \sim U(0, 1)$ regardless of what $\theta$ is. Under such a completely uninformative signal, a buyer that sees a seller without displayed quality is willing to pay for $E(\theta \mid \text{undisclosed}),$ whereas if he sees some seller with disclosed quality $s$ he is willing to pay for $$E(\theta \mid \text{disclosed}, s) = E(\theta \mid \text{disclosed})$$ as $s$ provides no information about the car's actual quality. Given this, a seller (of any type) will choose to disclose if $$E(\theta \mid \text{disclosed}) - c \geq E(\theta \mid \text{undisclosed})$$ Since this decision criteria is the same for all seller types (given the decisions of the other sellers to disclose or not), it follows that the sellers must either all disclose or all undisclose. The latter is not a reasonable equilibrium given that no information is conveyed to the buyer (as there is no nontrivial partition into disclose/undisclosed) and all sellers suffer a cost of $c,$ so we conclude that no types will disclose, as expected. Thus, the market failure is not prevented.

Although this simple analysis yields the trivial result that we expected - no one will pay for a completely uninformative signal - it does highlight the central incentive for a seller to disclose. There is a tradeoff between the cost incurred for disclosing by purchasing the DMV's certificate and the strategic ``cost" of having the quality hidden from the buyer. Intuitively, if the signal is informative, then higher types will disclose while lower types will not disclose - even if the signal is noisy. Thus, if a seller decides to not disclose the car quality, her perceived quality to the buyer may be lower as she will be grouped with other, lower quality sellers that also do not disclose. \newline

Now we turn to our central focus, the second signalling information structure, where the DMV charges $c$ and gives $\theta$ with probability $p > 0$ and with probability $1-p$ gives random $r \sim U(0,1)$ drawn uniformly, independent of $\theta.$ We say that this is the DMV's signal $s.$
\\
Then the buyer is willing to pay for:
\begin{itemize}
  \item $w = E(\theta \mid s = \emptyset) = E(\theta \mid \text{undisclosed}) = \widehat{\theta}$ if $\theta$ is not disclosed.
  \item $w = p \theta + (1-p) E(\theta \mid \text{disclosed})$ if $s = \theta$ is disclosed, because if the disclosed value is $\theta,$ with probability $p$ it is the true $\theta$ and with probability $1-p$ it is a fake signal and in that case the expected value of $\theta$ is $E(\theta \mid \text{disclosed}).$
  More formally, if the buyer observes signal $s,$ he is willing to pay for
  \begin{align*}
      E(\theta \mid s) &= \text{Pr} (\text{true signal}) E(\theta \mid s, \text{true signal}) + \text{Pr} (\text{fake signal}) E(\theta \mid s, \text{fake signal})  \\
      &= p \cdot s + (1 - p) \cdot E(\theta \mid \text{disclosed})
  \end{align*} 
  by total expectation.
\end{itemize} We also have by total expectation 
\begin{align*}
    E(\theta) = \frac{1}{2} &= \text{Pr} (\text{undisclosed}) E(\theta \mid \text{undisclosed}) + \text{Pr} (\text{disclosed}) E(\theta \mid \text{disclosed}) \\
    &= q \widehat{\theta} + (1-q) E(\theta \mid \text{disclosed})
\end{align*} 
And so $E(\theta \mid \text{disclosed}) = \frac{\frac{1}{2} - q \widehat{\theta}}{1 - q}$
where $q$ is the probability $\theta$ is not disclosed. \newline

Let $w(\theta)$ be the buyer's willingness to pay if he sees a seller with disclosed quality $\theta.$ Then we have  $$w(\theta) = p \theta + (1-p) E(\theta \mid \text{disclosed}) = p \theta + (1-p) \cdot \frac{ \frac{1}{2} - q \widehat{\theta}}{1-q}$$ 

Let $W(\theta_0)$ then be the expected profit (buyer's expected willingness to pay) for seller of type $\theta_0$ if she discloses. With probability $p$ the $\theta_0$ is accurately displayed to the buyer, in which case the buyer is willing to pay $w(\theta_0),$ and with probability $1-p$ some random $\theta \sim U(0,1)$ is drawn uniformly in which case the buyer is willing to pay $w(\theta).$ \\
Thus we have
\begin{align*}
    W(\theta_0) &= p w(\theta_0) + (1-p) \int_{0}^{1} w(\theta) \, d \theta \\
    &= p \left( p \theta_0 + (1-p) \cdot \frac{ \frac{1}{2} - q \widehat{\theta}}{1-q}  \right) + (1-p)  \int_{0}^{1} \left( p \theta + (1-p) \cdot \frac{ \frac{1}{2} - q \widehat{\theta}}{1-q} \right) \, d \theta
\end{align*} \\
So type $\theta_0$ will
\begin{itemize}
    \item disclose if $$W(\theta_0) - c > \widehat{\theta}$$
    \item not disclose if $$W(\theta_0) - c \leq \widehat{\theta}$$
\end{itemize}

Clearly $W(\cdot)$ is strictly increasing in $\theta_0$ as $\frac{\partial W}{\partial \theta_0}  = p^2 > 0$ \\
\newline
It follows that the car dealers still employ a cutoff strategy with $\theta \in [0,q]$ not disclosing, and so $\widehat{\theta} = \frac{q}{2},$ which means  $$E(\theta \mid \text{disclosed}) = \frac{\frac{1}{2} - q \widehat{\theta}}{1-q} = \frac{1+q}{2} = \frac{1+2\widehat{\theta}}{2}.$$ 
Thus 
\begin{align*}
    W(\theta_0) &= p \left(p \theta_0 + (1-p) \frac{1 + 2 \widehat{\theta}}{2} \right) + (1-p) \int_{0}^{1} \left( p \theta + (1-p) \frac{1 + 2 \widehat{\theta}}{2} \right) \, d \theta \\
    &= (1-p) \frac{1 + 2 \widehat{\theta}}{2} + p \cdot p \theta_0 + (1-p) \int_{0}^{1} p \theta \, d \theta \\
    &= (1-p) \frac{1 + 2 \widehat{\theta}}{2} + p^2 \theta_0 + (1-p) \frac{p}{2} \\
    &= (1-p) \frac{1 + p + 2 \widehat{\theta}}{2} + p^2 \theta_0
\end{align*}

Now solving: $$(1-p) \frac{1 + p + 2 \widehat{\theta}}{2} + p^2 \theta_0 - c > \widehat{\theta}$$

ie. 
$$ (1-p)(1 + p + 2 \widehat{\theta}) + 2 p^2 \theta_0 - 2c > 2 \widehat{\theta}$$
$$1 - p^2 + 2 p^2 \theta_0 - 2 c > p \cdot 2 \widehat{\theta}$$
$$2 p^2 \theta_0 > p \cdot 2 \widehat{\theta} + p^2 - 1 + 2c$$
$$\theta_0 > \frac{p \cdot 2 \widehat{\theta} + p^2 - 1 + 2c}{2 p^2}$$
\newline
So the types that don't disclose are:
$$\theta \leq \frac{p \cdot 2 \widehat{\theta} + p^2 - 1 + 2c}{2 p^2}$$ And so we have
$$\widehat{\theta} = E \left( \theta \, \, \middle\vert\, \,  \theta \leq \frac{p \cdot 2 \widehat{\theta} + p^2 - 1 + 2c}{2 p^2}  \right) = \min \left(    \frac{p \cdot 2 \widehat{\theta} + p^2 - 1 + 2c}{4p^2}       , \, \frac{1}{2}    \right)$$ Solving:
$$ 4 p^2 \widehat{\theta} = p \cdot 2 \widehat{\theta} + p^2 - 1 + 2c $$
$$2p(1 - 2p) \widehat{\theta} = 1 - p^2 - 2c$$
$$\widehat{\theta} = \min\left( \frac{1 - p^2 - 2c}{2p(1-2p)}, \, \frac{1}{2} \right)$$  We have the constraint: $$q = \frac{1 - p^2 - 2c}{p(1 - 2p)} \in [0,1]$$ And types that do disclose are $\theta_0 > 2 \widehat{\theta} = q$ Now we are maximizing:
$$ P = c(1 - 2 \widehat{\theta}) = c \left( 1 - \frac{1 - p^2 - 2c}{p(1-2p)}  \right) $$

Thus, the DMV solves the constrained optimization problem:

$$\max_{c, \, p} \hspace{0.1 cm} c \left( 1 - \frac{1 - p^2 - 2c}{p(1 - 2p)} \right)$$ subject to $$0 \leq \frac{1 - p^2 - 2c}{p(1 - 2p)} \leq 1, \hspace{0.5 cm} p \in [0,1]$$
We first consider $p > 1/2.$ Clearly this is a concave problem in $c$ for any fixed $p = \overline{p}$ as
$$c \left( 1 - \frac{ \overline{p}^2 - 2c}{\overline{p} (1 - 2 \overline{p})} \right) = c \left( 1 + \frac{\overline{p}^2}{\overline{p} ( 2 \overline{p} - 1)} - \frac{2c}{\overline{p} (2 \overline{p} - 1)}         \right)$$
where $\frac{ 1 }{ \overline{p} (2 \overline{p} - 1) } > 0$
so it is a downward opening sloping quadratic in $c.$ \\
Thus, the FOC (for fixed $p$) is

$$[c]: \hspace{0.5 cm}  1 - \frac{1 - p^2 - 2c}{p (1 - 2 p)} + c \left( \frac{2}{p(1 - 2p)}         \right) = 0$$
$$p(1 - 2p) - (1 - p^2) + 4c = 0$$
$$c = \frac{1 - p + p^2}{4}$$

Then profit is $$P^*(p) = \left( \frac{1 - p + p^2}{4} \right) \left(1 - \frac{1 - p^2 - 2 \cdot \left(  \frac{1 - p + p^2}{4}  \right) }{p(1 - 2p)}   \right)$$

And constraint is $$\frac{1 - p^2 - 2 \cdot \left(  \frac{1 - p + p^2}{4}  \right) }{p(1 - 2p)} \geq 0$$

$$ \frac{1 - p^2 - \tfrac{1}{2} (1 - p + p^2)}{p(1 - 2p)} = \frac{ 2 - 2p^2 - (1 - p + p^2)           }{2 p (1 - 2p) } = \frac{ 1 + p - 3p^2       }{2p(1 - 2p)} \geq 0$$ 

Thus, $1 + p - 3p^2 \leq 0 \implies p \geq p_{min} := \frac{1 + \sqrt{13}}{6} \approx 0.7676$

For $p > p_{min}$ then there is an interior solution.

At $p = p_{min}$ all types except lemons $(\theta = 0)$ will disclose.

We can simplify profit:
\begin{align*}
    P^*(p) &=  \left( \frac{1 - p + p^2}{4} \right) \left(1 - \frac{ 1 + p - 3p^2 }{ 2p(1 - 2p)   } \right) \\
    &= \left( \frac{1 - p + p^2}{4} \right) \left( \frac{p - 1 - p^2}{2p(1 - 2p)} \right) \\
    &=  \frac{ (1-p+p^2)^2 }{8p(2p - 1)} 
\end{align*}

As expected, when $p=1$ we have $P^*(1) = 1/8$ which is the baseline with no signal manipulation.

At $p = p_{min}$ the profit is:
$$P^* \left( \frac{1 + \sqrt{13}}{6} \right) =  \frac{11 - \sqrt{13}}{36} \approx 0.2054$$ Note that this already greatly exceeds the DMV's basline profit of $P^*(1) = 0.125$ under a perfect signal.
Now if $1/2 < p < p_{min}$ then the DMV chooses $c$ so that $\widehat{\theta} = 0$ is binding, so we have a corner solution, ie. all types except the lemons $(\theta = 0)$ disclose to maximize profit.

Thus we have 
\begin{align*}
    q = \frac{1 - p^2 - 2c}{p(1 - 2p)} &= 0 \\
    c = \frac{1 - p^2}{2}
\end{align*}
And we get $$P^*(p) = c \cdot (1 - 0) = \frac{1 - p^2}{2}$$
\newline
Now let us consider $p < 1/2.$ In this case, $P_p(c)$ is an upward facing parabola with minimum at $c = \frac{1 - p + p^2}{4}.$ And the constraint is $0 \leq \frac{1 - p^2 - 2c}{p (1 - 2p) } \leq 1$ and $1 - 2p > 0$ so rewrite this as 
$$0 \leq 1 - p^2 - 2c \leq p - 2p^2$$ which rearranges to 
$$ \frac{1 - p + p^2}{2} \leq  c \leq   \frac{1 - p^2}{2}$$

But note that $$\frac{1 - p + p^2}{4} \leq \frac{1 - p + p^2}{2} < \frac{1 - p^2}{2}$$ and thus clearly $P_p$ is maximized at $c = \frac{1 - p^2}{2}$ as $P_p$ is increasing on $( \frac{1 - p + p^2}{4}, \infty )$

Plugging this in we get $$P^*(p) = \frac{1 - p^2}{2} \left( 1 - \frac{1 - p^2 - 2 \cdot \frac{1 - p^2}{2}  }{p (1 - 2p)}   \right) = \frac{1 - p^2}{2}$$

And so for $p < 1/2$ the DMV again charges sufficiently low cost to induce all types except lemons $(\theta = 0)$ to disclose. \\

We can now summarize the optimal DMV profit $P^*(p)$ for all $p \in [0,1]$ as follows:

$$P^*(p) = \begin{dcases} \hspace{0.2 cm} \frac{(1 - p + p^2)^2}{8p(2p-1)}, & \hspace{0.3 cm} p \geq p_{min} \\ 
\hspace{0.2 cm} \frac{1 - p^2}{2}, &  \hspace{0.3 cm} 0 < p < p_{min} \\ \hspace{0.2 cm} 0, & \hspace{0.3 cm} p = 0 \end{dcases}$$
\newline

\begin{center}
    \includegraphics[width=12cm]{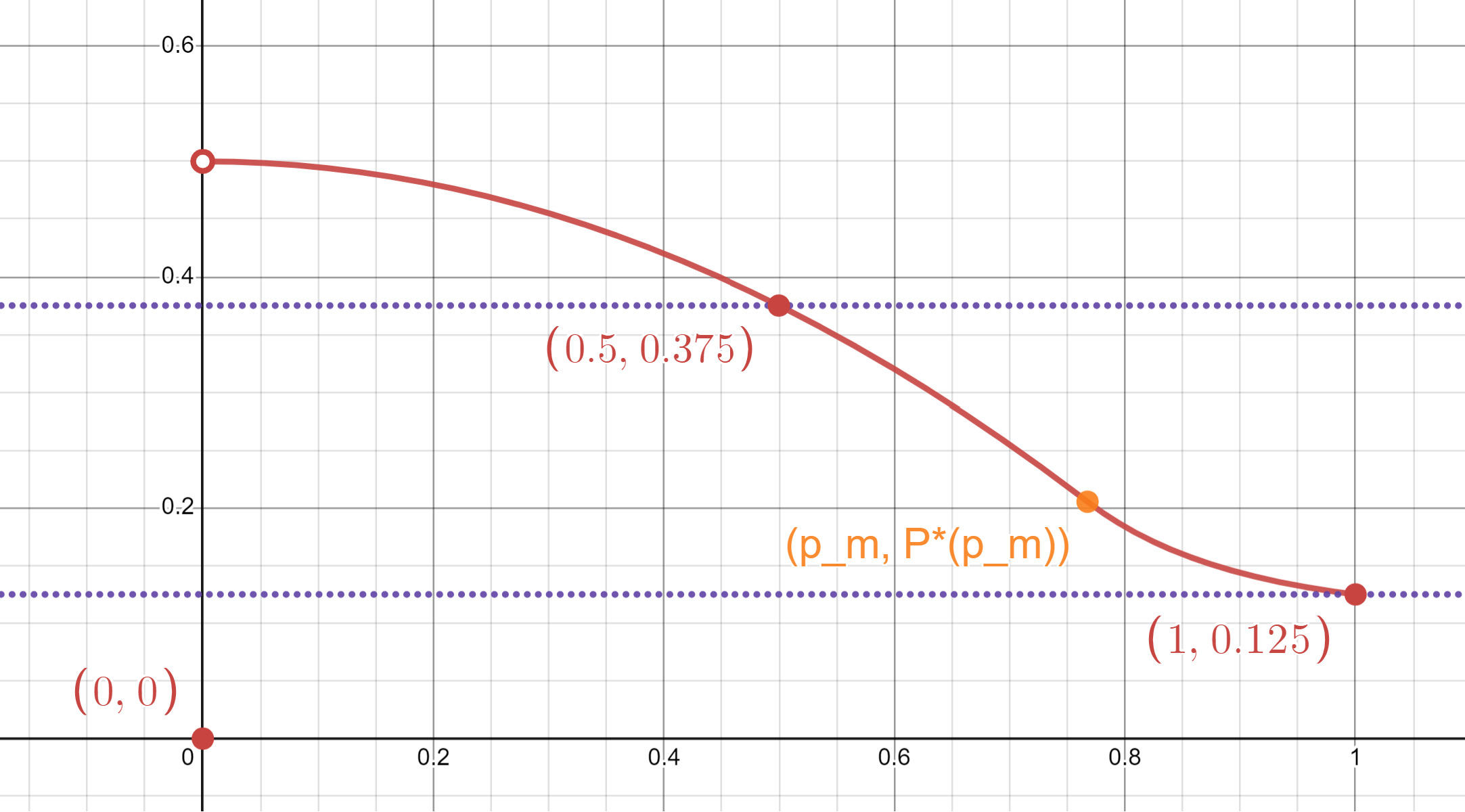}
\end{center}

The DMV's profit is increasing with decreasing $p,$ ie. by lowering the informativeness of the signal the DMV can make more overall. In fact the DMV can achieve any profit less than $(1 - 0^2) / 2 = 1/2$ by choosing sufficiently low values of $p.$ However, if the signal is completely uninformative ($p=0$) then the DMV makes no profit as no seller will purchase the certificate. Recall that in the baseline model with perfect information the DMV made profit $P = 1/8.$ \\

Thus the DMV can achieve almost \textbf{quadruple} the profit compared to the perfect signal in the baseline model. This is a surprising and compelling result. By degrading the quality of the signal, the DMV can actually increase its overall profit, even by charging the same amount $c$ per seller for any type. By introducing uncertainty into the signal, the DMV can actually encourage lower $\theta$ types to ``disclose," because there is some possibility that their revealed $s$ is higher than their true $\theta$ when the DMV gives the fake signal. Indeed, in the baseline model, only the sellers with types $\theta > 1/2$ disclose. And if $s \sim U(0, 1)$ then $E(s) = 1/2,$ so a seller of type $\theta < 1/2$ has an incentive to disclose with this possibility of misrepresenting its $\theta$ to a higher type. 

So by degrading the quality of the signal this way, the DMV incentivizes the lower type sellers to disclose and can charge more per seller. This continues until the DMV can incentivize all types except lemons ($\theta = 0$) to disclose, which maximizes profit.

We also have the optimal cost that the DMV charges,
$$c^*(p) = \begin{dcases} \hspace{0.2 cm} \frac{1 - p + p^2}{4}, & \hspace{0.3 cm} p \geq p_{min} \\ 
\hspace{0.2 cm} \frac{1 - p^2}{2}, &  \hspace{0.3 cm} 0 < p < p_{min} \\ \hspace{0.2 cm} 0, & \hspace{0.3 cm} p = 0 \end{dcases}$$

Note that $c^*$ is increasing on $(p_{min}, 1)$ and decreasing on $(0, p_{min}).$ \\ We see that for $p > p_{min},$ there is a tradeoff between lowering the cost per seller to attract more sellers (lower types) to disclose, and the profit made per seller that discloses. As $p$ decreases and the signal becomes less informative, lower types become willing to disclose, but the DMV continues to charge lower costs. At the threshold $p = p_{min},$ the DMV is able to capture the entire market at optimum, with all types except lemons disclosing. When the DMV further decreases $p,$ it can then begin to raise the cost again while still getting all non-lemons to disclose.

Perhaps the most surprising result is that even for $p < 1/2,$ the DMV continues to make more profit by further decreasing the quality of the signal, and charging even more per seller. For such $p,$ the DMV has $P^*(p) = c^*(p) = \frac{1 - p^2}{2}$ and all types except lemons ($\theta = 0$) disclose. This can be understood as follows: When the DMV continues to degrade the quality of its signal, this increases the lower types ($\theta < 1/2$) willingness to pay, because the probability of getting a fake signal, $1-p,$ increases, and the fake signal is on average better than the true $\theta$ being revealed. Thus, the lower types are wiling to pay more to receive the certificate. The higher types, on the other hand, suffer from the lower signal quality, but have no choice and still must disclose, because the alternative is to be grouped in with the undisclosed types, which are the very low types, ie. lemons. So they are willing to pay the higher cost of disclosing, even as the signal quality continues to degrade. This is the reason why the car dealers still employ a cutoff strategy for any $p > 0$ signal structure.

At optimum, the DMV charges cost $c$ equal to the lemons' reservation, just high enough to make the lemons indifferent and not disclose. Then the DMV is able to incentivize all higher types to disclose.

We observe, however, that by lowering the signal quality, overall welfare is clearly worse. The overall quality of disclosed $\theta$ is lower and buyers may be misled by fake signals. When $p < 1/2,$ even though every seller is on the market (which is pareto efficient), the DMV is cheating and giving a fake signal more than half the time.

In practice, it is unrealistic for the regulator to provide signals with such poor quality. However, in this model note that any nonzero value of $p < 1$ achieves strictly greater profit for the DMV than the profit from the baseline perfect signal. Even an almost perfect signal $p = 1 - \epsilon$ achieves higher profit than $p = 1.$ Thus, the DMV can exploit the information asymmetry in the market and increase its profit with imperfect signals.

\newpage

\textbf{Acknowledgements} \\

I would like to thank Profs. Scott Gehlbach and Zhaotian Luo at the UChicago Political Science department (and Political Economy program) for teaching me game theory and formal modelling in the social sciences in the SSI Formal Theory sequence during my first year. Prof. Luo discussed the classic Market for Lemons information asymmetry example with the baseline model (perfect signal) in his class and I am grateful to have discussed my generalizations with him. I also enjoyed seeing the Market for Lemons again this year in the ECON 201 class taught by Prof. Ivan Kwok.

\bigskip

\bigskip

\newpage
\textbf{References} \\

[1] $ $ Akerlof George (1970) The market for ‘Lemons’: quality uncertainty and the market mechanism. Quarterly Journal of Economics, 84:488–500

\end{document}